\documentclass[10pt,twocolumn,twoside]{IEEEtran}

\makeatother
\usepackage[english]{babel}
\usepackage[babel]{csquotes}
\usepackage{lipsum}
\usepackage{url}

\usepackage{emptypage}
\usepackage[english]{varioref}
\usepackage{color}
\usepackage{eurosym}
\usepackage{lettrine}

\usepackage{amsmath}
\usepackage{amsthm}
 \usepackage{multirow}
\usepackage{footmisc} 

\usepackage{mathrsfs}
\usepackage{amsfonts}
\usepackage{amssymb}
\newtheorem{theorem}{Theorem}
\newtheorem{proposition}[theorem]{Proposition}

\newtheorem{definition}{Definition}
\newtheorem{corollary}[theorem]{Corollary}

\newtheorem{remark}{Remark}
\newtheorem{assumption}{Assumption}
\newtheorem{example}{Example}
\usepackage{dsfont}
\usepackage{algpseudocode}
\usepackage{upgreek}

\usepackage{mathtools}

\usepackage{amscd}
\usepackage{booktabs}
\usepackage{graphics}
\usepackage{epsfig}
\usepackage{contour}
\usepackage{bm}
\usepackage{rotating}
\usepackage{tabularx}
\usepackage{longtable}
\usepackage{pdflscape}
\usepackage{cite}
\usepackage{listings}
\usepackage{xcolor}
\usepackage{siunitx}
\definecolor{green}{rgb}{0.13,0.55,0.13}

\newcommand{\red}[1]{{\textcolor{red}{#1}}}

\DeclareMathAlphabet{\mathpzc}{OT1}{pzc}{m}{it}
\DeclareMathAlphabet{\mathcal}{OMS}{cmsy}{m}{n}

\usepackage{cases}
\usepackage{tikz}
\usetikzlibrary{arrows,automata}
\usetikzlibrary{arrows,shapes,backgrounds,calc,positioning,patterns}
\usepackage{balance}
\usetikzlibrary{calc,arrows,shapes,backgrounds,calc,positioning,patterns,decorations.pathmorphing,decorations.markings,mindmap,trees}
\tikzstyle{block} = [draw, rectangle, minimum height=2em, minimum
width=4em] \tikzstyle{sum} = [draw, fill=blue!20, circle, node
distance=1cm] \tikzstyle{input} = [coordinate] \tikzstyle{output} =
[coordinate] \tikzstyle{pinstyle} = [pin edge={to-,thin,black}]
\usepackage{blox}
\usepackage{cases}
\usepackage{framed}
\colorlet{shadecolor}{black!15}
\usepackage{bigints}
\usepackage[american,cute inductors,smartlabels]{circuitikz}
\usetikzlibrary{arrows,shapes,backgrounds,calc,positioning,patterns}

\usepackage[american,cuteinductors,smartlabels]{circuitikz}

\usetikzlibrary{calc}
\ctikzset{bipoles/thickness=1} \ctikzset{bipoles/length=0.8cm}
\ctikzset{bipoles/diode/height=.375}
\ctikzset{bipoles/diode/width=.3}
\ctikzset{tripoles/thyristor/height=.8}
\ctikzset{tripoles/thyristor/width=1}
\ctikzset{bipoles/vsourceam/height/.initial=.7}
\ctikzset{bipoles/vsourceam/width/.initial=.7} \tikzstyle{every
node}=[font=\small] \tikzstyle{every path}=[line width=0.8pt,line
cap=round,line join=round]
\usepackage{flushend}

\newcommand{\kc}[1]{\textcolor{blue}{#1}}

\newcommand{\cE} { {\mathcal E}}
\newcommand{\cV} { {\mathcal V}}

\newcommand{\cB} { {\mathcal B}}
\newcommand{\bR} { {\mathbb R}}

\usepackage[ruled,vlined]{algorithm2e}

\usepackage{tikz}
\usetikzlibrary{arrows,automata}
\usetikzlibrary{arrows,shapes,backgrounds,calc,positioning,patterns}
\usepackage{balance}
\usetikzlibrary{calc,arrows,shapes,backgrounds,calc,positioning,patterns,decorations.pathmorphing,decorations.markings,mindmap,trees}
\definecolor{fgreen}{RGB}{204,223,181}
\definecolor{dblue}{RGB}{85,113,192}
\definecolor{dgreen}{RGB}{132,171,80} 
\usepackage[american,cute inductors,smartlabels]{circuitikz}
\usetikzlibrary{arrows,shapes,backgrounds,calc,positioning,patterns}

\usepackage[american,cuteinductors,smartlabels]{circuitikz}

\usetikzlibrary{calc}
\ctikzset{bipoles/thickness=1} \ctikzset{bipoles/length=0.8cm}
\ctikzset{bipoles/diode/height=.375}
\ctikzset{bipoles/diode/width=.3}
\ctikzset{tripoles/thyristor/height=.8}
\ctikzset{tripoles/thyristor/width=1}
\ctikzset{bipoles/vsourceam/height/.initial=.7}
\ctikzset{bipoles/vsourceam/width/.initial=.7} \tikzstyle{every
node}=[font=\small] \tikzstyle{every path}=[line width=0.8pt,line
cap=round,line join=round]

\makeatletter

\title{Reinforcement Learning based Distributed Control of Dissipative Networked Systems}
\author{ K. C. Kosaraju, S. Sivaranjani, W. Suttle, V. Gupta, and J. Liu 
\thanks{K.C. Kosaraju, V. Gupta are with the Department of Electrical Engineering, University of Notre Dame, Notre Dame, IN 46556, USA (email: \{kkosaraj, vgupta2\}@nd.edu).}
\thanks{S. Sivaranjani is with the Department of Electrical and Computer Engineering, Texas A\&M University, College Station, TX 77843, USA (email: sivaranjani@tamu.edu).}
\thanks{W. Suttle and J. Liu are with the Department of Applied Mathematics and Statistics and the Department of Electrical and Computer Engineering at Stony Brook University, Stony Brook, NY 11794, USA (email: \{wesley.suttle, ji.liu\}@stonybrook.edu).}
}
\date{\today}
\begin{document}
\maketitle
\begin{abstract}
We consider the problem of designing distributed controllers to stabilize a class of networked systems, where each subsystem is dissipative and designs a reinforcement learning based local controller to maximize an individual cumulative reward function. We develop an approach that enforces dissipativity conditions on these local controllers at each subsystem to guarantee stability of the entire networked system. The proposed approach is illustrated on a DC microgrid example, where the objective is maintain voltage stability of the network using local distributed controllers at each generation unit. 
\end{abstract}
\section{Introduction}
Distributed control of large scale networked systems is a classical research topic, with practical applications in a variety of fields such as transportation, chemical reaction, and hydraulic networks, multi-body mechanical systems, and  microgrids~\cite{egerstedt2001formation,sivaranjani2017distributed,dragivcevic2015dc,horn1972general,lasseter2004microgrid}. The problem provides many challenges such as non-classical information patterns, computational complexity due to the large state-space, scalability of control design methods, complex system dynamics that may be imperfectly known, and so on. Despite many important advances, the field continues to be a focus of intense research.

An interesting direction in recent times has been the utilization of reinforcement learning for distributed and multi-agent control. Reinforcement Learning (RL) is especially powerful for the control of systems where the dynamics and/or the environment are unknown \cite{sutton2018reinforcement}. In a typical RL-based design, the aim is to learn a controller that maximizes its cumulative reward while exploring the unknown environment. A wide variety of model-based and model-free algorithms are now available (see, e.g., \cite{kaelbling1996reinforcement} for a survey). While initially developed for single agent settings, the scope of RL based techniques has also been expanded to multi-agent networked systems (see \cite{busoniu2008comprehensive,zhang2019multi,zhang2019decentralized} for surveys). Further, while the typical focus of RL-based techniques for controller design has been through simulations and demonstrations, a growing line of research now considers obtaining guarantees about concerns traditional to control theory, e.g., stability, safety, and robustness, through controllers obtained using RL~\cite{cheng2019end}.

In this paper, we consider the problem of guaranteeing stability when RL is used for distributed control of networked dynamical systems. Specifically, consider a large scale system consisting of many subsystems that are coupled through their inputs and outputs, such as a network of microgrids. Each subsystem designs a local controller based on information about the subsystem state, inputs, and outputs. In particular, we assume that the controller is implemented using an RL algorithm since the dynamics of the subsystems may be unknown. Of note, however, different controllers may potentially use different RL algorithms. How do we design the controllers that guarantee that the entire system is still stable? There are at least two challenges here. First, we would like the control strategy to be distributed.  While there exists a wide literature on RL techniques for multi-agent systems, distributed control strategies using RL that provide guarantees like stability, safety, and robustness \cite{bucsoniu2018reinforcement} are still scant. Works that consider the problem of guaranteeing stability and robustness with RL controllers have largely been limited to contexts such as model-based RL and LQR designs for  single-agent systems \cite{lewis2012reinforcement,berkenkamp2017safe,fazel2018global,zhang2019policy}. 
Second, most available literature on multi-agent RL considers the case when all subsystems implement the same RL algorithm and further share information such as a global state or rewards with other subsystems. Development of RL-based controllers at the subsystems that ensure stability and robustness for the entire  networked system, especially when different agents may not use the same RL algorithm, largely remains an open problem.

As a first step towards addressing this problem, we focus on a class of networked systems where each subsystem is dissipative~\cite{willems1972dissipative} in open loop. Dissipativity is an input-output concept that can be used to guarantee a broad range of useful properties such as $\mathcal{L}_2$ stability, robustness with respect to disturbances, and stability under time-delays \cite{van2000l2, desoer2009feedback, niemeyer1991stable} and has been widely used in traditional control theory for distributed controller synthesis~\cite{chopra2006passivity, arcak2016networks,van2013port, agarwal2019compositional,agarwal2020distributed, 9093201,tippett2013dissipativity, sivaranjani2018mixed,agarwal2019sequential}. In the context of RL, dissipativity has been used to enhance the convergence/performance of various learning schemes \cite{gao2020passivity} and has been enforced as a system property for specific systems like Port-Hamiltonian systems \cite{nageshrao2014passivity,sprangers2014reinforcement}. However, there has been limited literature on enforcing it using model-free RL techniques or on exploring its potential to permit distributed controller design that guarantees properties such as stability at the system level. The challenge in our formulation is that an RL controller aiming to optimize the local performance metric at a subsystem can easily disrupt the dissipativity of the subsystem with respect to the variables that it exchanges with the other subsystems.  

In this paper, we develop a reinforcement learning based distributed control design approach that exploits the dissipativity property of individual subsystems to guarantee stability of the entire networked system. Our proposed approach can be summarized as follows. We first 
use a control barrier function to characterize the set of controllers that enforce a dissipativity condition at each subsystem (Propositions~\ref{prop:dissipativity_using_cbf} and \ref{thm:dissipativity_RL}). 
We impose a minimal energy perturbation on the control input learned by the RL algorithm to project it to an input in this set (Theorem~\ref{thm:dissipativity_RL}). Together, these results guarantee the stability of the entire networked system even when the subsystems utilize potentially heterogeneous RL algorithms to design their local controllers (Theorem~\ref{prop:stab_of_network}). 

Our approach of utilizing a control barrier function (CBF) to impose the constraint that the controller designed for each subsystem using RL preserves the dissipativity of the subsystem in the closed loop parallels the use of CBFs to enforce safety in RL algorithms \cite{cheng2019end}. CBFs guarantee the existence of control inputs under which a super-level set of a function (typically representing specifications like safety) is forward invariant under a given dynamics~\cite{ames2016control,romdlony2014uniting,wieland2007constructive}. However, their use to impose input-output properties such as dissipativity is less studied. Here, we utilize CBFs to characterize the set of dissipativity ensuring controllers, and then learn a dissipativity ensuring controller for each subsystem from this set.

The main contribution of this work is a distributed approach to ensure stability of a networked system  with dissipative subsystems when the individual subsystems utilize RL to design their own controllers. Beyond the specific stabilization problem that we focus on, integrating dissipativity (and other input-output) specifications into RL-based control is useful since it allows a wide landscape of tools from classical dissipativity theory to be integrated into RL-based control design. The proposed algorithm guarantees stability irrespective of the choice of the RL algorithm used at each subsystem. In particular, the results also hold for heterogeneous RL algorithms being used at each subsystem. We also note that as opposed to most existing literature on multi-agent RL, the proposed approach requires only the output from neighboring subsystems to learn the control policy at each subsystem. In other words, to guarantee stability, no information about the states, rewards, or policies of other subsystems is required. 

The paper is organized as follows. In Section \ref{sec:network_sys}, we present the model of the networked system, state the necessary assumptions, and provide the problem formulation.  In Section~\ref{sec:controlbf}, we utilize CBFs to characterize the set of controllers that guarantees dissipativity of each subsystem. In Section~\ref{sec:RL}, we present an RL algorithm to compute a control input  that preserves the dissipativity of each subsystem, and show that it stabilizes the networked system. In Section \ref{sec:case_study}, we numerically illustrate our approach on a Direct-Current microgrid application. Finally, in Section \ref{sec:conc}, we provide some directions for future work. Proofs of all the results in the paper, and the definitions of dissipativity, are provided in the Appendix.

\textbf{Notation:} $\mathbb{R}^{m}$ denotes the space of $m$-dimensional real vectors, $\mathbb{R}$ denotes the space of real numbers, and $\bR_+$ denotes the set of all positive real numbers. $\otimes$ denotes the Kronecker product. $z^{\top}$ denotes the transpose of a vector or a matrix $z$ and $\|z\|_{2}$ (or simply $\|z\|$) denotes its 2-norm. For a symmetric matrix $M$ and a vector $z$ of compatible dimensions, $\|z\|^2_{M}$ is defined to be equal to $z^{\top}Mz.$ Given square matrices $M_{1}$, $M_{2}$, $\cdots$, $M_{n}$, define the matrix $\textrm{diag}(M_{i})$ as the block diagonal matrix whose main-diagonal blocks are matrices $M_{1}$, $M_{2}$, $\cdots$, $M_{n}$, and all off-diagonal blocks are zero matrices. For a symmetric matrix $M$, $\lambda_{min}(M)$ denotes its smallest eigenvalue. $I$ denotes the identity matrix with dimensions clear from the context. A directed graph $\mathcal{G}=\left(\mathcal{V}, \mathcal{E}\right)$ is defined by a finite set of nodes (or vertices) $\mathcal{V}$ and a set of directed edges (or arcs) $\mathcal{E},$ together with a mapping from $\mathcal{E}$ to the set of pairs of $\mathcal{V}$. By convention, we disregard self-loops. Thus, to any arc $e\in \cE$, there corresponds an ordered pair $(u,v)\in \cV\times \cV,$ with $u\neq v,$ representing the head vertex $u$ and the tail vertex $v$. Given this, a shorthand notation is to simply say $(u,v)\in \cE.$ A graph is undirected if whenever $(u,v)\in \cE$ then $(v,u)\in \cE$. The in-neighbor set $\mathcal{N}_{i}$ of node $i$ is the set of all vertices $j$ such that $(j,i)\in \cE.$ Let $\mathcal{D} \subset \bR^n$. A function $f:\mathcal{D}\rightarrow \bR^n$ is Lipschitz if there exists a constant $L$ satisfying $\|f(b)-f(a)\|_2 \leq L\|b-a\|_2$ for all $a,~b \in \mathcal{D}$, and class $C^{1}$ if it is continuously differentiable. We denote a value obtained by sampling the probability distribution function $f_{X}(x)$ for a random variable $X$ as $y\sim f_{X}(x).$ When the random variable is clear from the context, we denote the distribution function simply by $f(x).$ 

\begin{figure}
    \centering
    \includegraphics[trim=0cm 0cm 0cm 0cm, clip=true, width=\columnwidth]{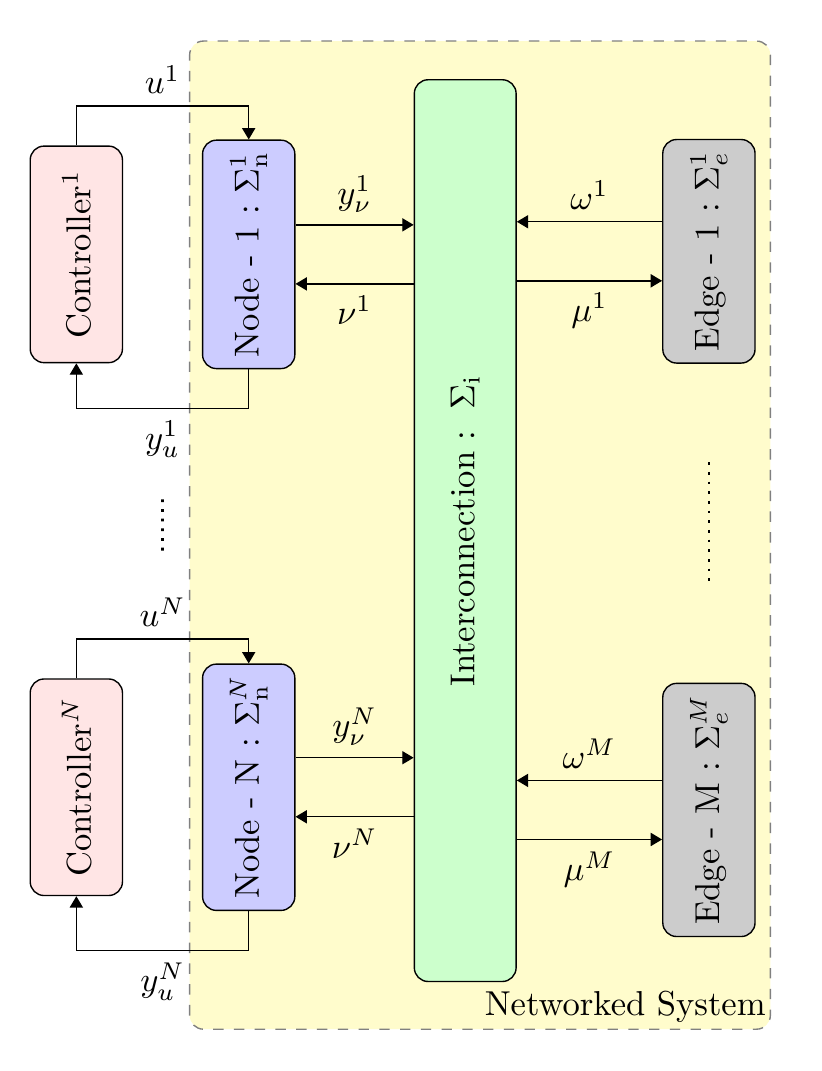}
    \caption{Schematic of the system configuration}
    \label{fig:config}
\end{figure}
\section{Problem Formulation}\label{sec:network_sys}
We adapt the general framework described in~\cite{van2013port} and shown in Figure~\ref{fig:config}. 

\textbf{Node dynamics:} Consider a networked system described by a directed graph $\mathcal{G}=(\mathcal{V},\mathcal{E})$, where each node $i \in V$ is a subsystem $\Sigma^i_n$, given by
\begin{align}\label{sys:network:node_dynamics}
\begin{split}
\Sigma_{\mathrm{n}}^i:~\left\{
\begin{matrix}
x_{t+1}^i\\
y^i_{u,t}\\
y^{i}_{\nu,t}
\end{matrix} 
\begin{matrix}
=\\
=\\
=
\end{matrix}
\begin{matrix}
f^i(x^i_t, u^i_t, \nu^i_t)\\
g^i(x^i_t, u^i_t)\\
h^i(x^i_t, \nu^i_t)
\end{matrix}
\right.
\end{split}
 \end{align}
where at time $t$, $x^i_t\in \mathbb{R}^{n_{i}}$ denotes the state of the $i$-th subsystem,  $u^i_t\in \mathbb{R}^{m_{i}}$ denotes the control input applied by the subsystem controller that needs to be designed, and $\nu^i_t\in \mathbb{R}^{p_{i}}$ is the input to the $i$-th subsystem that depends on the output of the other subsystems in the in-neighbor set of node $i$. The subsystem has two outputs:  $y_{u, t}^{i}\in\mathbb{R}^{\overline{o}_i}$ which is the output that is used to design the control input $u_{t}^{i}$, and $y_{\nu, t}^{i}\in\mathbb{R}^{\hat{o}_i}$ which is the output that is used to compute the inputs  $\nu^j_t$ for other subsystems $j$ for whom $i$ is an in-neighbor. We will define the exact relation between $\nu^i_t$ and $y_{t}^{j}$, $j\in\mathcal{N}_{i}$, later. Given that each subsystem corresponds to a unique node in the graph, we use the terms subsystem dynamics and node dynamics interchangeably. We assume that the state transition function $f^i$ and the output functions $g^i,~h^i$ are of Class $C^1$. Without loss of generality we assume that $(x^i=0, u^i=0, \nu^i=0)$ is an equilibrium point of the subsystem $\Sigma_{\mathrm{n}}^i$.

For future reference, define $x^\top \triangleq [x^{1\top},\ldots,x^{N\top}]\in \bR^n$, $u^\top \triangleq [u^{1\top},\ldots,u^{N\top}]\in \bR^m$, $y^\top_u \triangleq [y^{1\top}_u,\ldots,y^{N\top}_u]\in \bR^{\overline o}$, $y^\top_{\nu} \triangleq [y^{1\top}_{\nu},\ldots,y^{N\top}_{\nu}]\in \bR^{\hat{o}}$, $y^i\triangleq [y^{i\top}_{u},y^{i\top}_{\nu}]\in \bR^{o}$, $y^\top \triangleq [y^{1\top},\ldots,y^{N\top}]\in \bR^o$, and $\nu^\top \triangleq [\nu^{1\top},\ldots,\nu^{N\top}]\in \bR^p$.

As stated earlier, definitions of dissipativity are provided in Appendix \ref{sec:dissipativity} for the sake of completeness. We make the following assumption throughout the paper.
\begin{assumption}[Dissipative  node dynamics]\label{ass:open_loop_dissipativity}
Each subsystem $\Sigma_{n}^{i}$ with dynamics defined in \eqref{sys:network:node_dynamics} is dissipative, in the set $\mathcal{S}^i_n$, with respect to the supply function 
\begin{align}\label{network:open_loop_dissipativity}
\begin{split}
    w^i_n(u^i, \nu^i, y_{u}^i, y_{\nu}^i) = & \underbrace{u^{i\top}S^{i\top}_uy_{u}^i  - \|u^i\|^2_{R_u^i}  - \|y_{u}^i\|^2_{Q_u^i}}_{\triangleq w^i_{u}(u^i, y^i_u)}\\
    &+\underbrace{\nu^{i\top}S^{i\top}_{\nu}y_{\nu}^i - \|\nu^i\|^2_{R^i_{\nu}} -\|y_{\nu}^i\|^2_{Q^i_{\nu}}}_{\triangleq w^i_{\nu}(\nu^i,y^i_{\nu})},
\end{split}
\end{align}
where $S^{i}_u,~R_u^i=\left(R_{u}^{i}\right)^{\top},~Q_u^i=\left(Q_{u}^{i}\right)^{\top},$ $S^{i}_{\nu},~R^i_{\nu}=\left(R_{\nu}^{i}\right)^{\top},$ and $Q^i_{\nu}=\left(Q_{\nu}^{i}\right)^{\top}$ are matrices of appropriate dimensions.
\end{assumption}
For future reference, define $S_u\triangleq\textrm{diag}(S_{u}^{i})$, $R_u\triangleq\textrm{diag}(R_{u}^{i})$, $Q_u\triangleq\textrm{diag}(Q_{u}^{i})$,
$S_{\nu}\triangleq\textrm{diag}(S_{\nu}^{i})$, $R_{\nu}\triangleq\textrm{diag}(R_{\nu}^{i})$, and $Q_{\nu}\triangleq\textrm{diag}(Q_{\nu}^{i})$.
Further, denote $\epsilon_{\nu} = \lambda_{min}\left(Q_{\nu}\right)$, $\delta_\nu = \lambda_{min}\left(R_{\nu}\right)$, $\epsilon_u = \lambda_{min}\left(Q_u\right)$, $\delta_u = \lambda_{min}\left(R_u\right)$, $\epsilon_{e} = \lambda_{min}\left(Q_{e}\right)$ and $\delta_e = \lambda_{min}\left(R_{e}\right)$.
\begin{remark}\label{rem:no_diss_clp}
Even though Assumption~\ref{ass:open_loop_dissipativity} states that the subsystem is dissipative, it is an assumption in the `open loop'. Note that the design of the controller that determines the inputs $u^{i}$ has not been specified. The dissipativity property required for system stability concerns the inputs $\mu^{i}$ and the outputs $y_{\mu}^{i}$ and this may easily be disrupted by the additional dynamics, say of the form $u^{i}=\zeta^i(x^i)$, introduced through the design of the controller. For a simple illustration of this fact, note that from \cite[Corollary 4.1.5]{l2gain}, Assumption \ref{ass:open_loop_dissipativity}  holds if and only if the condition
\begin{align}
    \sum_{t=t_0}^{t-1}\sum_{i=1}^{N}\left(w^i_{u}(u^i_t, y^i_{u_t}) + w^i_{\nu}(\nu^i,y^i_{\nu})\right) \geq 0,
\end{align}
holds for all $0\leq t_0\leq t$. Consider subsystem \eqref{sys:network:node_dynamics} in closed-loop with a Lipschitz controller $u^i=\zeta^i(x^i) \in \bR^{m_i}$. Then, we notice that
\begin{align}
    \sum_{t=t_0}^{t-1}\sum_{i=1}^{N} w^i_{\nu}(\nu^i_t,y^i_{\nu, t}) \geq -\sum_{t=t_0}^{t-1}\sum_{i=1}^{N}w^i_{u}(\zeta^i(x^i_t), y^i_{u,t}) \ngeq 0,
\end{align}
which implies that unless the controller has been designed to ensure that $w^i_{u}(\zeta^i(x^i), y^i_{u})\leq 0$, dissipativity of the subsystem in the closed loop with the controller may not be preserved.
\end{remark}
\textbf{Edge dynamics:} While the simplest form of coupling among the subsystems would be to equate the inputs $\nu^{i}_{t}$ for the  subsystem $i$ with the output $y_{t}^{j}$ of subsystem $j$ if $(j,i)\in \mathcal{E}$, inspired by~\cite{van2013port}, we consider a more general model that allows the edges in the graph $\mathcal{G}$ to be described a dynamic system as well. Specifically for edge $k\in\cE$, the dynamics are given by
\begin{align}\label{sys:network:edge-dynamics}
\begin{split}
\Sigma_{\mathrm{e}}^{k}:~\left\{
\begin{matrix}
z_{t+1}^{k}\\
\omega^{k}_t
\end{matrix} 
\begin{matrix}
=\\
=
\end{matrix}
\begin{matrix}
g^k(z_t^k, \mu_t^k)\\
j^k(z_t^k, \mu_t^k)
\end{matrix}
\right.
\end{split}
 \end{align}
where $z^{k}_t \in \bR^{q_i}$ denotes the edge subsystem state at time $t$,  $\mu^{k}_t\in \bR^{r_i}$ denotes the input at time $t$, and $\omega^{k}_t\in \bR^{s_i}$ denotes the output at time $t$. We assume that the state transition function $g^k$ and the output function $j^k$ are of Class $C^1$. Once again, without loss of generality we assume that $(z^k=0, \mu^k=0)$ is an equilibrium point of the subsystem $\Sigma_{\mathrm{e}}^k$. For future reference, define $z^\top \triangleq [z^{1\top},\ldots,z^{M\top}]\in \bR^q$, $\omega^\top \triangleq [\omega^{1\top},\ldots,\omega^{M\top}]\in \bR^s$, and $\mu^\top \triangleq [\mu^{1\top},\ldots,\mu^{M\top}]\in \bR^r$, where $M$ denotes the cardinality of the set $\cE$. 
\begin{assumption}[Dissipative  edge dynamics]\label{ass:open_loop_dissipativity_edge}
Each subsystem $\Sigma_e^k$ with its dynamics defined in \eqref{sys:network:edge-dynamics} is dissipative in the set $\mathcal{S}_{e}^{k}$ with supply-function 
\begin{align}\label{network:open_loop_dissipativity_edge}
    w^k_e(\mu^k, \omega^k) = & \mu^{k\top}S^{k\top}_{e}\omega^k  - \|\mu^k\|^2_{R^{k}_{e}}  - \|\omega^k\|^2_{Q^{k}_{e}},
\end{align}
where $S^{k}_e,~R^{k}_{e}=\left(R^{k}_{e}\right)^{\top},~Q^{k}_{e}=\left(Q^{k}_{e}\right)^{\top}$ are matrices of appropriate dimensions.
\end{assumption}
For future reference, define $S_e\triangleq\textrm{diag}(S_{e}^{k}$), $R_e\triangleq\textrm{diag}(R_{e}^{k}$), and $Q_e\triangleq\textrm{diag}(Q_{e}^{k})$.

\textbf{Interconnection among subsystems:} The entire networked system is defined through the interconnection of the subsystems defined by the nodes and edges by relating the inputs $\nu$ and outputs $y_{\nu}$ of the node subsystems with the inputs $\mu$ and outputs $\omega$ of the edge subsystems as specified below. Define $s^\top \triangleq [x^\top, z^\top]$ as the state variable of the overall network. Further, define
\begin{align}\label{eq:dissi[ativity_modified}
\begin{split}
        w_u(u, y_u)&\triangleq \left(u^{\top}S^{\top}_uy_{u}  - \|u\|^2_{R_{u}}  - \|y_{u}\|_{Q_{u}}^2 \right),\\
    w_{\nu}(\nu, y_{\nu})&\triangleq\left(\nu^{\top}S^{\top}_{\nu}y_{\nu} - \|\nu\|_{R_{\nu}}^2  -\|y_{\nu}\|_{Q_{\nu}}^2\right),\\
    w_e(\mu, \omega)& \triangleq\left( \mu^{\top}S^{\top}_e\omega - \|\mu\|_{R_{e}}^2  - \|\omega\|_{Q_{e}}^2\right).
\end{split}
\end{align}
Following~\cite{van2013port}, we model the interconnection among the subsystems through the equation
 \begin{align}\label{sys:graph:inter_laws_gen}
\Sigma_{\mathrm i}:~     \begin{bmatrix}
     \nu\\
     \mu
     \end{bmatrix} = 
     \begin{bmatrix}
     0 & \mathcal{B}\\
     -\mathcal{B}^\top & 0
     \end{bmatrix}
     \begin{bmatrix}
     y_{\nu}\\
     \omega
     \end{bmatrix}
 \end{align}
for a suitably defined matrix $\mathcal{B}$.  Further,  we make the following assumption.
\begin{assumption}\label{ass:S}
Matrices $S_{\nu}$ and $S_e$ in \eqref{eq:dissi[ativity_modified} satisfy
\begin{align}\label{eqn:S}
    \mathcal B^\top S_{\nu}^\top - S_e\mathcal{B}^{\top} = 0
\end{align}
\end{assumption}
An interpretation of~\eqref{sys:graph:inter_laws_gen} and Assumption~\ref{ass:S} is that the edges of the system do not generate any energy.  Although equation~(\ref{sys:graph:inter_laws_gen}) appears intricate, most interconnected physical systems can be written in this form (see~\cite{van2013port} for examples from various domains; an example of interconnected distributed generation units is discussed in detail below). Similarly, several relevant subclasses  of  dissipative  systems  including,  but  not limited  to,  $\mathcal{L}_2$ gain systems and passive systems satisfy Assumption~\ref{ass:S}, see \cite{arcak2016networks} for other examples.
For future reference, denote  \begin{align}
\label{eq:b_definition}
    \mathcal{B}_{\delta}(x)&\triangleq\epsilon_eI + x\mathcal{B}^\top \mathcal{B},\\
    \mathcal{B}_{\epsilon}(y)&\triangleq y I + \delta_e \mathcal{B} \mathcal{B}^\top.
\end{align}
   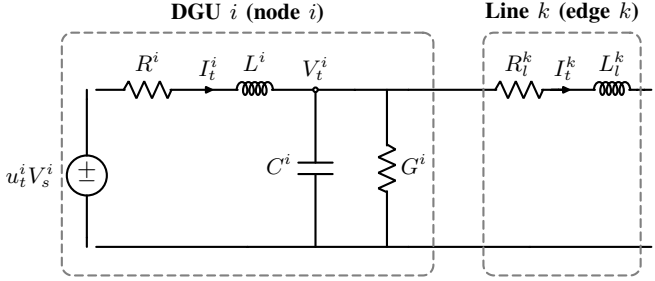
\begin{figure}[t]
    	\begin{center}
    		\begin{circuitikz}[scale=.95,transform shape]
    			\ctikzset{current/distance=1}
    			\draw
    			node[] (Ti) at (0,0) {}
    			node[] (Tj) at ($(5.4,0)$) {}
    			node[] (Aibattery) at ([xshift=-4.5cm,yshift=0.9cm]Ti) {}
    			node[] (Bibattery) at ([xshift=-4.5cm,yshift=-0.9cm]Ti) {}
    			node[] (Ai) at ($(Aibattery)+(0,0.2)$) {}
    			node[] (Bi) at ($(Bibattery)+(0,-0.2)$) {}
    			(Ai) to [R, l={$R^i$}] ($(Ai)+(1.7,0)$) {}
    			($(Ai)+(1.7,0)$) to [short,i={$I^i_t$}]($(Ai)+(1.701,0)$){}
    			($(Ai)+(1.701,0)$) to [L, l={$L^i$}] ($(Ai)+(3,0)$){}
    			to [short, l={}]($(Ti)+(0,1.1)$){}
    			(Bi) to [short] ($(Ti)+(0,-1.1)$);
    			\draw
    			($(Ai)$) to []($(Aibattery)+(0,0)$)to [V_=$u^i_t V_s^i$]($(Bi)$)
    			($(Ti)+(-1.3,1.1)$) node[anchor=south]{{$V_{t}^i$}}
    			($(Ti)+(-1.3,1.1)$) node[ocirc](PCCi){}
    			($(Ti)+(-.3,1.1)$) to [R, l={$G^{i}$}] ($(Ti)+(-.3,-1.1)$)
    			($(Ti)+(-1.3,1.1)$) to [C, l_={$C^i$}] ($(Ti)+(-1.3,-1.1)$)
    			($(Ti)+(2.,1.1)$) to [short,i={$I_{t}^k$}] ($(Ti)+(2.2,1.1)$)
    			($(Ti)+(0,1.1)$)--($(Ti)+(.6,1.1)$) to [R, l={$R^{k}_l$}] 
    			($(Ti)+(2.5,1.1)$) {} to [L, l={{$L^k_l$}}, color=black]($(Tj)+(-2.2,1.1)$){}
    			($(Tj)+(-2.2,1.1)$) to [short]  ($(Ti)+(3.4,1.1)$)
    			($(Ti)+(0,-1.1)$) to [short] ($(Ti)+(3.4,-1.1)$);
    			\draw
    			node [rectangle,draw,minimum width=5.2cm,minimum height=3.4cm,dashed,color=gray,label=\textbf{DGU $i$ (node $i$)},densely dashed, rounded corners] (DGUi) at ($0.5*(Aibattery)+0.5*(Bibattery)+(2.25,0.2)$) {}
    			node [rectangle,draw,minimum width=2.2cm,minimum height=3.4cm,dashed,color=gray,label=\textbf{Line $k$ (edge $k$)},densely dashed, rounded corners] (DGUi) at ($0.5*(Aibattery)+0.5*(Bibattery)+(6.65,0.2)$) {};
    		\end{circuitikz}
    		\caption{Electrical scheme of DGU $i$ and transmission line $k$ as considered in Example~\ref{ex:dc_microgrid}.}
    		\label{fig:networks}
    	\end{center}
    	\vspace{.2cm}
    \end{figure}
    \begin{figure}[t]
\centering
\begin{tikzpicture}[scale=0.8,transform shape,->,>=stealth',shorten >=1pt,auto,node distance=3cm,
                    semithick]
  \tikzstyle{every state}=[circle,thick,fill=white,draw=black,text=black]

  \node[state] (A)                    {\num{1}};
  \node[state]         (B) [above right of=A] {\num{2}};
  \node[state]         (D) [below right of=A] {\num{4}};
  \node[state]         (C) [below right of=B] {\num{3}};

  \path (A)  edge   [below] node {\hspace{7mm}$I_{l,t}^1$} (B) 
           (B) edge      [below]        node {\hspace{-7mm}$I_{l,t}^2$} (C)
           (C) edge         [above left]     node {$I_{l,t}^3$} (D)
           (D) edge 	[above right]     node {$I_{l,t}^4$} (A);
\end{tikzpicture}
\caption{The topology of network considered in Example~\ref{ex:dc_microgrid}.}\label{fig:microgrid_example}
\end{figure}
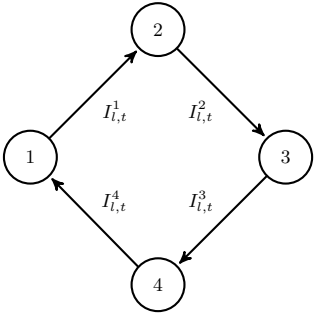
\begin{example}\label{ex:dc_microgrid}
Consider  the electrical schematic of a microgrid, containing four Distributed Generating Units (DGUs) and interconnected through four transmission lines, as shown in Figures \ref{fig:networks} and \ref{fig:microgrid_example}. The DGUs correspond to the nodes  and the transmission lines correspond to the edges of the graph describing this networked system. Let the DGUs and the transmission lines be numbered as shown in Fig~\ref{fig:microgrid_example}. Each DGU contains a DC-DC buck converter that is operating on a constant impedance load. The controller to be designed sets $u_{t}^{i}\in(0,1)$ for the $i$-th DGU. Denote by $I_{t}^{k}$ the current through the $k$-th transmission line at time $t$ and by $V_{t}^{i}$ the voltage across the $i$-th DGU at time $t$. Define the state of the subsystem at the $i$-th node (corresponding to the $i$-th DGU) by $x_{t}^{i}\triangleq \left[\begin{array}{lr}I_{t}^{i}&V_{t}^{i}\end{array}\right]^{\top}.$ The dynamics of the $DGU$ at node $i \in \mathcal{V} :=  \left\{1\ldots 4\right\}$, which forms the $i$-th subsystem, can be written as
\begin{align}\label{dyn:node_microgrid}
\begin{split}
    I_{t+1}^i &= I_t^i - (T_s/L^i)(R^iI^i_t + V^i_t - u^i_tV_s)\\
    V_{t+1}^i&= V_t^i + (T_s/C^{i})(I^i_t - G^iV^i_t + \nu^i_t), 
    \end{split}
\end{align}
where $T_{s}, L^i,~C^i,~R^i,~G^i, ~V_s^i \in \bR_{>0}$ are constants, $u^i_{t}\in (0,~1)$ is the local control input to be designed, and $\nu_{t}^{i}\in\bR$ is the input to the $i$-th subsystem that depends on the output of the other subsystems in its in-neighbor set through the relations
\begin{equation}
    \left[\begin{array}{c}\nu_{t}^{1}\\\nu_{t}^{2}\\\nu_{t}^{3}\\\nu_{t}^{4}\end{array}\right]=\left[\begin{array}{c}I_{l, t}^4-I_{l, t}^1\\
           I_{l, t}^1-I_{l, t}^2\\
           I_{l, t}^2-I_{l, t}^3\\
           I_{l, t}^3-I_{l, t}^4\end{array}\right],
\end{equation}
where $I_{l, t}^k$ denotes the current through the edge $k$. We denote the outputs $y_{\nu,t}^{i}\triangleq V_{t}^{i}.$

The edges correspond to the transmission lines connected to each DGU. The dynamics of the transmission line at edge $k \in \mathcal{E} := \left\{1\ldots 4\right\}$ are given by
\begin{align}\label{dyn:edge_microgrid}
\begin{split}
         I_{l,t+1}^k &= I_{l,t}^k - (T_s/L_{l}^k)(R_{l}^k I_{l,t}^k + \mu^k_t)\\
    \omega_t^k &= I_{l,t}^k
\end{split}
\end{align}
where $L_l^k,~R_l^k \in \bR_{>0}$ are constants, $I_{l,t}^k\in \bR$ denotes the state variable, and $\mu^k_{t}\in \bR$ denotes the input from the nodes connected to the edge $k$ defined as
\begin{equation}
    \left[\begin{array}{c}\mu_{t}^{1}\\\mu_{t}^{2}\\\mu_{t}^{3}\\\mu_{t}^{4}\end{array}\right]=\left[\begin{array}{c}V_t^2-V_t^1\\
           V_t^3-V_t^2\\
           V_t^4-V_t^3\\
           V_t^1-V_t^4\end{array}\right].
\end{equation}
Define the incidence matrix $\mathcal{B} \in \bR^{4 \times 4}$ to model the network topology. Specifically, if the ends of each edge $k$ are arbitrarily labeled with a $+$ and a $-$, then the entries of  $\cB$ are given by
    \begin{equation*}
    \cB_{ik}=
    \begin{cases}
    +1 \quad &\text{if $i$ is the positive end of $k$}\\
    -1 \quad &\text{if $i$ is the negative end of $k$}\\
    0 \quad &\text{otherwise}.
    \end{cases}
    \end{equation*}
    The interconnection between the nodes and edges can then be expressed as
 \begin{align}
\begin{bmatrix}
     \nu_t\\
     \mu_t
     \end{bmatrix} = 
     \begin{bmatrix}
     0 & \mathcal{B}\\
     -\mathcal{B}^\top & 0
     \end{bmatrix}
     \begin{bmatrix}
     y_{\nu,t}\\
     \omega_t
     \end{bmatrix} = \begin{bmatrix} \mathcal{B} I_{l,t}\\ \mathcal{B}^\top V_t\end{bmatrix}
 \end{align}
\end{example}

\textbf{Controller design:} We assume that each subsystem $i$ wishes to design its controller to maximize the expected discounted cumulative reward, 
\begin{align}\label{exp:J_pi}
    J^i = \mathbb{E}\left[\sum_{t=0}^{\infty}\gamma^t r^i_{t}(x_t^i,u_t^i)\right],
\end{align}
where $\gamma \in (0,1)$ is the discount factor,   $r^i_{t}(x_t^i,u_t^i)$ is the per step reward function evaluated at time $t$, and the expectation is over any stochasticity that may arise due to the control policy itself. We assume that each agent utilizes a RL algorithm to design its controller. For a given control policy $\pi^i$, we define the value function $V_{\pi}^i$, and the state-action value function $Q_{\pi}^i$ below:
\begin{align}\label{exp:V_pi}
    V_{\pi}^i(x^i) &= \mathbb{E}_{\pi^i} \left[\sum_{t=0}^{\infty} \gamma^t r^i_{t}(x^i_t, u^i_t) \ | \ x^i_0 = x^i \right], \\
    Q^i_{\pi}(x^i, u^i) &= \mathbb{E}_{\pi^i} \left[ \sum_{t=0}^{\infty} \gamma^t r^{i}_{t}(x^i_t, u^i_t) \ | \ x^i_0 = x^i, u^i_0 = u^i \right],
    \label{exp:Q_pi}\\
    A_{\pi}^i(x^i, u^i)&=Q^i_{\pi}(x^i, u^i)-V_{\pi}^i(x^i).\label{exp:advantage_fun}
\end{align} 
Note that we do not assume that each subsystem utilizes the same RL algorithm. However, we assume that the RL algorithms converge.

\textbf{Problem statement:} Equations~\eqref{sys:network:node_dynamics}, \eqref{sys:network:edge-dynamics} and \eqref{sys:graph:inter_laws_gen} jointly define the networked system $\Sigma$ under consideration, with state defined as $s_{t}^{\top}\triangleq\left[x_{t}^{\top},z_{t}^{\top}\right].$ From Assumption \ref{ass:open_loop_dissipativity}, we know that the each subsystem $i$ is dissipative with the supply-function $w^i_u(u^i, y^i_u) + w^i_{\nu}(\nu^i, y^i_{\nu})$.  However, since the subsystems use RL to design their local controllers, the closed loop subsystems may not remain dissipative (see Remark \ref{rem:no_diss_clp}). Further, the control actions of all the subsystems may end up destabilizing the entire networked system. We are interested in the problem of how to design the RL algorithm at each subsystem to guarantee the stability of the networked system. Specifically, consider a networked system on a directed graph $\mathcal{G}=(\mathcal{V},\mathcal{E})$, described by \eqref{sys:network:node_dynamics}, \eqref{sys:network:edge-dynamics}, and \eqref{sys:graph:inter_laws_gen}, and satisfying Assumptions \ref {ass:open_loop_dissipativity}, \ref{ass:open_loop_dissipativity_edge} and \ref{ass:S}. Assume that the controller at each subsystem $i$ is designed using an RL algorithm to maximize the discounted cumulative reward $J^i$ in~(\ref{exp:J_pi}). How should the updates in the RL algorithms be done so that the control policies at convergence guarantee Lyapunov stability of the overall networked system?


\section{Dissipativity ensuring Reinforcement Learning} \label{sec:diss_pres_rl}
In this section, we present the main results of the paper through a new distributed RL algorithm that guarantees the stability of the entire networked system. The proposed approach is as follows.
\begin{itemize}
    \item[(a)] {\bf Control barrier functions for dissipativity}: As stated in {Remark \ref{rem:no_diss_clp}}, even though each subsystem $i$ is dissipative with supply-function $w^i_u(u^i, y^i_u) + w^i_{\nu}(\nu^i, y^i_{\nu}),$ with the controller for the input $u^{i}$ the subsystem may no longer remain {dissipative with the input-output pair $ w^i_{\nu}(\nu^i, y^i_{\nu})$}. Our first step is to utilize control barrier functions to characterize the set of all controllers that ensure that the closed loop subsystem $i$ is dissipative with respect to the input $\nu^{i}$ and output $y^{i}$ (c.f. Fig~\ref{fig:config}) with the supply function
\begin{align}\label{desired:supply-function}
    w_d^i(\nu^i, y^i) &= \nu^{i\top}S_{\nu}^{i\top}y_{\nu}^i - \delta^i_{d}\|\nu^i\|^2_2 -\epsilon^i_{d}\|y^i\|^2_2,
\end{align}
where $\delta^i_{d}\in \bR$ and $\epsilon^i_{d} \in \bR$ are tuning parameters set by the designer. 
    \item[(b)]{\bf Projection-based RL algorithm for dissipativity}:  In the second step, at each subsystem $i$, we consider the control input generated by an RL algorithm that seeks to maximize the discounted cumulative reward given by \eqref{exp:J_pi} and use a quadratic program (QP) to project this control input onto the set of control inputs that ensure that the closed loop subsystem remains dissipative with supply-function $w^i_d(\nu^i, y^i_{\nu})$. Note that the RL algorithms used at different nodes can be different. 
    \item[(c)]{\bf Networked system stability}: We finally show that if each subsystem designs the controller to ensure that it is dissipative, the entire networked system is also stable. 
\end{itemize}

We now develop these steps one by one. We will make the following assumption in the sequel.
\begin{assumption}
\label{ass:set}
Denote $\alpha=\min(\delta_{d}^1,\ldots,\delta_{d}^N)$, and $\beta=\min(\epsilon_{d}^1,\ldots, \epsilon_{d}^N)$. The conditions 
\begin{align}\label{eq_ass:des_alpha_beta}
\begin{split}
    \mathcal{B}_{\delta}(\alpha) &\geq 0,\\
    \mathcal{B}_{\epsilon}(\beta)&\geq 0,
\end{split}
\end{align} 
hold, where $\mathcal{B}_{\delta},$ and $\mathcal{B}_{\epsilon}$ have been defined in~(\ref{eq:b_definition}).
\end{assumption}

\subsection{Control barrier functions for dissipativity}
\label{sec:controlbf}
Control barrier functions (CBFs) are now a popular tool for enforcing safety constraints in nonlinear control systems. The following definition follows the development in~\cite{ames2019control, xu2015robustness, notomista2020persistification}. 
\begin{definition}[Time-varying Zeroing Control Barrier Functions] \label{def:cbf}
Consider a function $b:\mathbb{R}_+\times \mathbb{R}^{n+q} \rightarrow \mathbb{R}$ that is continuously differentiable in both arguments. Define a closed set $\mathcal{C}$ as the super-level set of this function as follows:
\begin{align}\label{set:forward_inv}
    \mathcal{C} \triangleq \left\{s\in \bR^{n+q} \;| \;b(t,s)\geq 0\right\}.
\end{align}
The function $b(t,s_t)$ is a time-varying zeroing control barrier function, for the networked system $\Sigma$ described by \eqref{sys:network:node_dynamics}, \eqref{sys:network:edge-dynamics} and \eqref{sys:graph:inter_laws_gen} and with state $s_{t}$, if there exists an $\eta \in [0,~1]$ such that for all $s_t\in \mathcal C$, $t\in \mathbb{R}_+$,
\begin{align}\label{barrier_suff_cond:gen}
  \sup_{u_t \in \bR^m} \left[b(t+1, s_{t+1})+(\eta-1)b(t, s_t)\right] \geq 0.\\\nonumber
\end{align}
\end{definition}
Control barrier functions can be used to derive sufficient conditions under which a super-level set of a function of the state of the networked system $\Sigma$ is forward invariant. These conditions also characterize the set of control inputs achieving such forward invariance through the relation
\begin{align}\label{ctrl:deterministic_pol}
    \mathrm{B}(t,s_t) \triangleq \left\{u_t\in \bR^m|b(t+1, s_{t+1})+(\eta-1)b(t, s_t)\geq 0\right\}.
\end{align}
The following result, given for completeness for a discrete time setting such as ours, shows that the set $\mathcal{C}$ defined in \eqref{set:forward_inv} is forward invariant for every $u_t\in \mathrm{B}(t,s_t)$. 
\begin{proposition}[Discrete-time time-varying Control Barrier Functions]\label{prop:cbf} 
Consider a time-varying zeroing control barrier function $b(t,s_t)$ and its super level set $\mathcal{C}$ defined in \eqref{set:forward_inv}. Then any Lipschitz input $u_t\in \mathrm{B}(t,s_t)$, where $\mathrm{B}(t,s_t)$ is given in \eqref{ctrl:deterministic_pol}, will render the set $\mathcal{C}$ forward invariant. 
\end{proposition}

Although dissipativity is a property defined by the input, and the output, we can utilize control barrier functions to characterize the set of controllers that ensures dissipativity in the closed loop of the subsystems, which in turn guarantee the stability of the overall networked system \cite{notomista2019passivity} . Following Proposition \ref{prop:cbf}, we define a control barrier function for each subsystem $i$ as follows. Denote 
\begin{align}\label{desired:diss_error}
    \tilde{w}^i(u^i, \nu^i, y_{u}^i, y_{\nu}^i) \triangleq w_n^i(u^i, \nu^i, y_{u}^i, y_{\nu}^i)-w_d^i(\nu^i, y^i).
\end{align}
Then, define the control barrier function \begin{equation}
    b^i(t,x^i_t) \triangleq -\sum_{\tau =t_0}^{t-1}\tilde w^i(u^i_{\tau}, \nu^i_{\tau}, y_{u_,\tau}^i, y_{\nu,\tau}^i),
\end{equation} 
whose super-level set is given by
\begin{align}\label{cbf:single_node}
    \mathcal C^i = \left\{x^i_t\in \bR^{n_i} \;| \; b^i(t,x^i_t) \geq 0\right\}.
\end{align}
To use the control barrier function $b^i(t,x^i_t)$ to enforce dissipativity of the closed loop subsystem, we proceed as follows. Denote
	\begin{multline}\label{policy:dissipativity_barrier}
	\mathrm{D}^i(x^i_t, \nu_t^i) \triangleq \{u^i\in \bR^{m_i}|-\tilde w^i(u^i_{t}, \nu^i_{t}, y_{u,t}^i, y_{\nu,t}^i)\\+\eta^i b^i(t, x^i_t)\geq 0\}, 
	\end{multline}
where $\eta^i \in [0,~1]$ is a designer specified parameter. 
We can then state the following result.
\begin{proposition}[Control barrier function for dissipativity] \label{prop:dissipativity_using_cbf}
	Consider the problem formulation in Section~\ref{sec:network_sys}. If $u^{i}\in \mathrm{D}^i(x^i, \nu^i)$ at all time steps, then the subsystem~\eqref{sys:network:node_dynamics} is dissipative with respect to input $\nu^{i}$ and output $y^{i}$ with supply function $ w^i_{d}(\nu^i,y^i_{\nu})$.
\end{proposition}
From Proposition \ref{prop:dissipativity_using_cbf}, if the set $\mathrm{D}^i(x^i_t, \nu_t^i)$ is non-empty, then any control input $u_t^i \in \mathrm{D}^i(x^i_t, \nu_t^i)$  renders \eqref{sys:network:node_dynamics} dissipative with respective to the supply function $ w^i_{d}(\nu_t^i,y_{\nu_t}^i)$. We can choose a particular control input in this set from other considerations, such as minimizing the control cost. We can also use this set to ensure that the control input from an RL algorithm ensures that the subsystem is dissipative as shown next. 

\subsection{Dissipativity ensuring RL policies}
\label{sec:RL}
We now consider the case when an RL algorithm is used for designing the control inputs $u^{i}$ and show how the input can be chosen to one that preserves the dissipativity of the closed-loop subsystem $\Sigma_{n}^{i}$ with respective to the supply function $ w^i_{d}(\nu^i,y_{\nu}^i)$. The key idea is similar to shielded RL techniques \cite{cheng2019end,alshiekh2018safe,fisac2018general} and uses the control barrier function based characterization of the set of dissipativity ensuring controllers obtained above to both project the control policy and to guide the future exploration of the RL algorithm. 

We assume that the RL algorithm proceeds in an episodic fashion. Let $\pi^{\mathrm{RL}_i}_{k}$ denote the policy at the $k$-th policy iteration of the RL algorithm. This policy will in general be stochastic and may be parameterized by some parameters $\theta_k^i$ that may correspond to, e.g., the neural network being used to learn the policy. The paramterization is not relevant to our arguments and to minimize notational complexity, we suppress it in the sequel. Let $u^{\mathrm{RL_i}}_{k}(x_t^i)\sim \pi^{\mathrm{RL}_i}_{k}(\cdot|x_t^i)$. Our algorithm proceeds by projecting this input on the set of dissipativity ensuring controllers. Specifically, we propose that the overall dissipativity ensuring control input in the $k$-th episode takes the following structure:
\begin{align}\label{policy:compensated}
	u_k^{\mathrm{DEC}_i}(x_t^i) &= u_k^{\mathrm{FF}_i}(x^i_t) + u_k^{\mathrm{CBF}_i}(x^i_t, u_k^{\mathrm{FF}_i}),
\end{align}
where 
$u_k^{\mathrm{FF}_i}(x^i)$ 
represents the feedforward compensation, given by
\begin{align}\label{policy:uncompensated}
u_k^{\mathrm{FF}_i}(x^i_{t}) &= u^{\mathrm{RL_i}}_{k}(x^i_{t}) + \sum_{j=0}^{k-1}u_j^{\mathrm{CBF}_i}(x^i_{t},u_j^{\mathrm{FF}_i}(x^{i}_{t})),
\end{align}
and $u_k^{\mathrm{CBF}_i}$ is computed using the optimization problem:
\begin{align}\label{QP:RL_diss}
u_k^{\mathrm{CBF}_i}(x^i_t, u_k^{\mathrm{FF}_i}) = & \arg\min_{a_t^i \in \bR^{m_i}}\|a_t^i\|\\
\nonumber\text{s.t.} &   -\tilde w(u_t^i, \nu_t^i, y_{u_t}^i, y_{\nu}^i)+\eta^i b^i(t, x^i_t) \geq 0, \\
\nonumber& a_t^i+u_{k}^{\mathrm{FF}_i}(x^i_t) = u_t^i. 
\end{align}
As in the usual control barrier function based works, the formulation in the relation~(\ref{policy:compensated}) seeks to minimize the energy of the perturbation needed to project the control input in the set of dissipativity ensuring controllers \cite{cheng2019end, ames2019control}. The feedforward compensation in~(\ref{policy:uncompensated}) is split into two parts: $u_{k}^{\mathrm{RL}_i}(x^i)$ represents the control input obtained from the RL policy. However, this might not ensure dissipativity of the closed loop subsystem. The second term in~(\ref{policy:uncompensated}) represents our best guess to rectify the input to ensure dissipativity. 
Furthermore, the term $u^{\mathrm{CBF}_i}$ in \eqref{policy:compensated} may be interpreted as the  feedback part of the controller. The complete algorithm description is given in Algorithm~\ref{algo:DEC}.

\begin{algorithm}\label{algo:DEC}
	\SetAlgoLined
	\For{$i = 1,\ldots, N $}{
	Initialize RL input $\pi_0^{\mathrm{RL}_i}$, and arrays $\hat{D}^i$ and $\hat{A}^i$.
		}
	\For{$t = 0,\ldots, T $}{
	    \For{$i = 1,\ldots, N $}{
	    Sample $u^{\mathrm{RL}_{i}}_{0}(x_{t}^{i})\sim\pi^{\mathrm{RL}_i}_{0}$ and compute $u_0^{\mathrm{CBF}_i}(x_{t}^{i},u_{0}^{FF_{i}})$ using  \eqref{QP:RL_diss}.\\ 
		Deploy $u_0^i(x^{i}_{t}) = u^{\mathrm{RL}_{i}}_{0}(x^{i}_{t}) + u_0^{\mathrm{CBF}_i}(x_{t}^{i},u_{0}^{FF_{i}})$\\
		Store state-action pairs $(x_t^i, u_0^{\mathrm{CBF}_i}(x_{t}^{i},u_{0}^{FF_{i}}))$ in $\hat{A}^i$\\
		}
		\For{$i = 1,\ldots, N $}{Observe $x^i_t, u_0^i(x^{i}_{t}), x_{t+1}^i, r_t^i$ and store in $\hat{D}^i$ for use in the RL algorithm
		}
	}
	\For{$i = 1,\ldots, N $}{Collect Episode Reward $\sum_{t=1}^Tr_t^i$
		}
	
	Set $k=1$ (representing the $k$-{th} episode or input iteration step)
	
	\While{$k<$ Max\_Episodes}{
	\For{$i = 1,\ldots, N $}{Do input iteration using RL algorithm based on previously observed episode to obtain $\pi^{\mathrm{RL}_i}_{k}$
	
		}
		Initialize state $s_0$ from an initial state distribution
		
		\For{t = 0,\ldots, T }{
		\For{$i = 1,\ldots, N $}{
			Compute the feed-forward term $u_k^{\mathrm{FF}_i}(x_{t}^{i}) = u^{\mathrm{RL_i}}_{k}(x_{t}^{i}) + \sum_{j=0}^{k-1}u_j^{\mathrm{CBF}_i}(x^{i}_{t},u_{j}^{FF_{i}}(x_{t}^{i}))$
			
			Use \eqref{QP:RL_diss} solve for $u_k^{\mathrm{CBF}_i}(x^{i}_{t},u_{k}^{FF_{i}})$ 
			
			Deploy controller $u_k^i(x^{i}_{t})= u_k^{\mathrm{FF}_i}(x_t^i) + u_k^{\mathrm{CBF}_i}(x^i_t, u_k^{\mathrm{FF}_i}(x_t^i))$
	
		Store state-action pairs $(x_t^i, u_k^i(x^{i}_{t}))$
		
		}
		\For{$i = 1,\ldots, N $}{Observe $x^i_t, u_k^i(x^{i}_{t}), x_{t+1}^i, r_t^i$ and store in $\hat{D}^i$ for use in the RL algorithm
		}
	}
	$k=k+1$
	}
	\caption{RL-DEC algorithm.}
\end{algorithm}
We assume that the parameter $Max\_Episodes$ has been chosen to be large enough that the algorithm converges. Upon convergence, denote $u^{\textrm{DEC}_{i}}(x_{t}^{i})$ to be the final deployed controller $u_{k}^{i}(x_{t}^{i})$ for $k=Max\_Episodes$. The following result shows that Algorithm~\ref{algo:DEC} renders the closed loop subsystem dissipative. For brevity, we skip the proof as it is a direct consequence of Proposition 2 and Definition 2.
\begin{proposition}\label{thm:dissipativity_RL}
Consider the problem formulation in Section~\ref{sec:network_sys}. Let the controller $u^{\textrm{DEC}_{i}}(x_{t}^{i})$ designed with Algorithm~\ref{algo:DEC} be used as the input $u_{t}^{i}$ for the subsystem \eqref{sys:network:node_dynamics}. If there exists a solution to the optimization problem \eqref{QP:RL_diss} for all $(x^i, \nu^i)$, then the closed-loop subsystem \eqref{policy:compensated} is dissipative with supply function $ w^i_{d}(\nu^i,y_{\nu}^i)$.
\end{proposition}

\begin{remark}
Computing $u_k^{\mathrm{FF}_i}(x)$ requires the solution of the optimization problem $k$ times; further, the knowledge of all $u^{\mathrm{RL_i}}_{0}, \ldots, u^{\mathrm{RL}_i}_{{k-1}}$ is required. Consequently, for large $k$, the proposed algorithm can become memory intensive and computationally expensive. However, we need not compute $u_k^{\mathrm{FF}_i}(x)$ very accurately because of the presence of the feedback term $u_{k}^{CBF_{i}}$. This raises the possibility of  approximating $u_k^{\mathrm{FF}_i}(x)$ by using a feed-forward neural network  $u^{\mathrm{bar}}_{\phi_k}$ to learn the term $ \sum_{j=0}^{k-1}u_j^{\mathrm{CBF}_i} $. In this case, \eqref{policy:uncompensated} should be replaced by 
\begin{align}\label{policy:uncompensated_approx}
u_k^{\mathrm{FF}_i}(x) &= u^{\mathrm{RL_i}}_{k}(x) + u^{\mathrm{bar}}_{\phi_k^i}(x),
\end{align}
where $\phi_k$ parameterizes the neural network, which is updated using the data from previously collected samples.  \end{remark}

The following is the main result of the paper, which shows that the controller  calculated using Algorithm~\ref{algo:DEC} stabilizes the networked system. 
\begin{theorem}[Stability of networked system in closed-loop]\label{prop:stab_of_network}
Consider the problem formulation in Section \ref{sec:network_sys} with Assumption~\ref{ass:set}. If $u_t^i$ is chosen to be equal to $u^{\mathrm{DEC}_i}(x_t^i)$ at all time steps and for all subsystems $i$, then the  networked system defined by \eqref{sys:network:node_dynamics}, \eqref{sys:network:edge-dynamics} and \eqref{sys:graph:inter_laws_gen} is Lyapunov stable with respect to the origin. Further, suppose that $\mathcal{B}_{\delta}(\alpha) > 0$, $\mathcal{B}_{\epsilon}(\beta)> 0$, and $R_{u}\triangleq\textrm{diag}(R_{u}^{i})>0$. If the systems \eqref{sys:network:node_dynamics}, and \eqref{sys:network:edge-dynamics} are zero state detectable, then the networked system defined by \eqref{sys:network:node_dynamics}, \eqref{sys:network:edge-dynamics}, and \eqref{sys:graph:inter_laws_gen} is also asymptotically stable with respect to the origin.
\end{theorem}
The definition of {\em zero-state detectability} is provided in Definition 3 of Appendix A.
\begin{remark}[Decentralized and Distributed]
In \eqref{QP:RL_diss}, each agent needs to evaluate $\tilde{w}$ which requires the information of $\nu_t$. From \eqref{sys:graph:inter_laws_gen}, computing $\nu_t$ requires  information from its neighbours. Then, the proposed RL algorithm is distributed.  However, in the event when the desired supply-function $w_d$ is equal to $w_{\nu}$, then $\tilde{w} = w_u$. Consequently, the RL algorithm takes a decentralized form.
\end{remark}

\section{Case study: DC Microgrid}\label{sec:case_study}
We now evaluate the proposed control barrier function based RL Algorithm \ref{algo:DEC} in simulation. We consider the DC microgrid in Example \ref{ex:dc_microgrid} with 4 DGU's, interconnected through resistive and inductive lines as shown in Figure \ref{fig:microgrid_example}. The control objective is to regulate the voltage $V^i$ across the load of each DGU's to its desired value $\overline{V}^i \in \bR$.
Thus, we define the set of all feasible forced equiliria of the node subsystems \eqref{dyn:node_microgrid} and the edge  subsystems \eqref{dyn:edge_microgrid} as
\begin{align}
 \mathcal{C}_{i}^n = \left\{(\overline I^i,\overline V^i,\overline u^i,\overline \nu^i) \in \mathbb{R}^{4}|\right.&\left.R^i\overline{I}^i+\overline{V}^i - \overline{u}^iV_s^i=0,\right. \\&\left.\overline{I}^i - G\overline{V}^i + \overline{\nu}^i=0\right\}, \nonumber
\end{align}
and
\begin{align}
    \mathcal{C}_{k}^e = \left\{(\overline I_l^i,\overline \mu^i) \in \mathbb{R}^{2}|R_l^i\overline{I}_l^i+\overline{\mu}^i=0\right\},
\end{align}
respectively. 
{In the development above, we have assumed that $(s=0)$ is the desired equilibrium. However, the results are agnostic to the choice of the equilibrium. Since the objective in this case study is to stabilize the system at a non-trivial operating point $(\overline I^i,\overline V^i,\overline u^i,\overline \nu^i, \overline I_l^i,\overline \mu^i)\in \mathcal{C}_{i}^n \times \mathcal{C}_{k}^e$, 
%
we shift the equilibrium of the networked system to the trivial equilibrium via a simple change of variables. In what follows, for a given variable $\nu$, denote the error between $\tilde{\nu} = \nu - \overline{\nu}$.}

In \cite{cucuzzella2019voltage}, the authors show that the subsystems at the node \eqref{dyn:node_microgrid} and the edge   \eqref{dyn:edge_microgrid} are dissipative with the supply-functions 
\begin{align}\label{network:microgrid_node_dissipativity}
\begin{split}
    w^i_n(\tilde u^i, \tilde \nu^i, \tilde y_{u}^i, \tilde y_{\nu}^i) = & \underbrace{\tilde u^{i\top}\tilde y_{u}^i  - R^i\|\tilde y_{u}^i\|^2_2}_{w^i_{u}(\tilde u^i, \tilde y^i_u)}+\underbrace{\tilde \nu^{i\top}\tilde y_{\nu}^i  -G^i\|\tilde y_{\nu}^i\|^2_2}_{w^i_{\nu}(\tilde \nu^i,\tilde y^i_{\nu})}
\end{split}
\end{align}
and
\begin{align}\label{network:microgrid_edge_dissipativity}
\begin{split}
   w^k_e(\tilde \mu^k, \tilde \omega^k) = & \tilde \mu^{k\top}\tilde \omega^k   - R^k_l\|\tilde \omega^k\|^2_2,
\end{split}
\end{align}
respectively. As a next step, we define the desired supply function corresponding to {\eqref{desired:supply-function}} as
\begin{align*}
    w_d^i(\tilde\nu^i, \tilde y^i) = w^i_{\nu}(\tilde\nu^i,\tilde y^i_{\nu})- R^i\|\tilde y_{u}^i\|^2_2
\end{align*}
where we chose $\delta_d^i = 0,~\epsilon_d^i = R^i$, which satisfies equation \eqref{eq_ass:des_alpha_beta} in Assumption \ref{ass:set}. Consequently, using \eqref{desired:diss_error} we  compute the  resulting control barrier function as 
\begin{align}
    b^i(t,x^i_t) = -\sum_{\tau =t_0}^{t-1}\left(\tilde u^{i\top}\tilde y_{u}^i-G^i\|\tilde y_{\nu}^i\|^2_2\right), \; t\geq t_0\geq0,
\end{align}
and its super-level is defined as in \eqref{cbf:single_node}.
%
%
\begin{figure}
    \centering
    \includegraphics[trim=0cm 0cm 0cm 0cm, clip=true, width=\columnwidth]{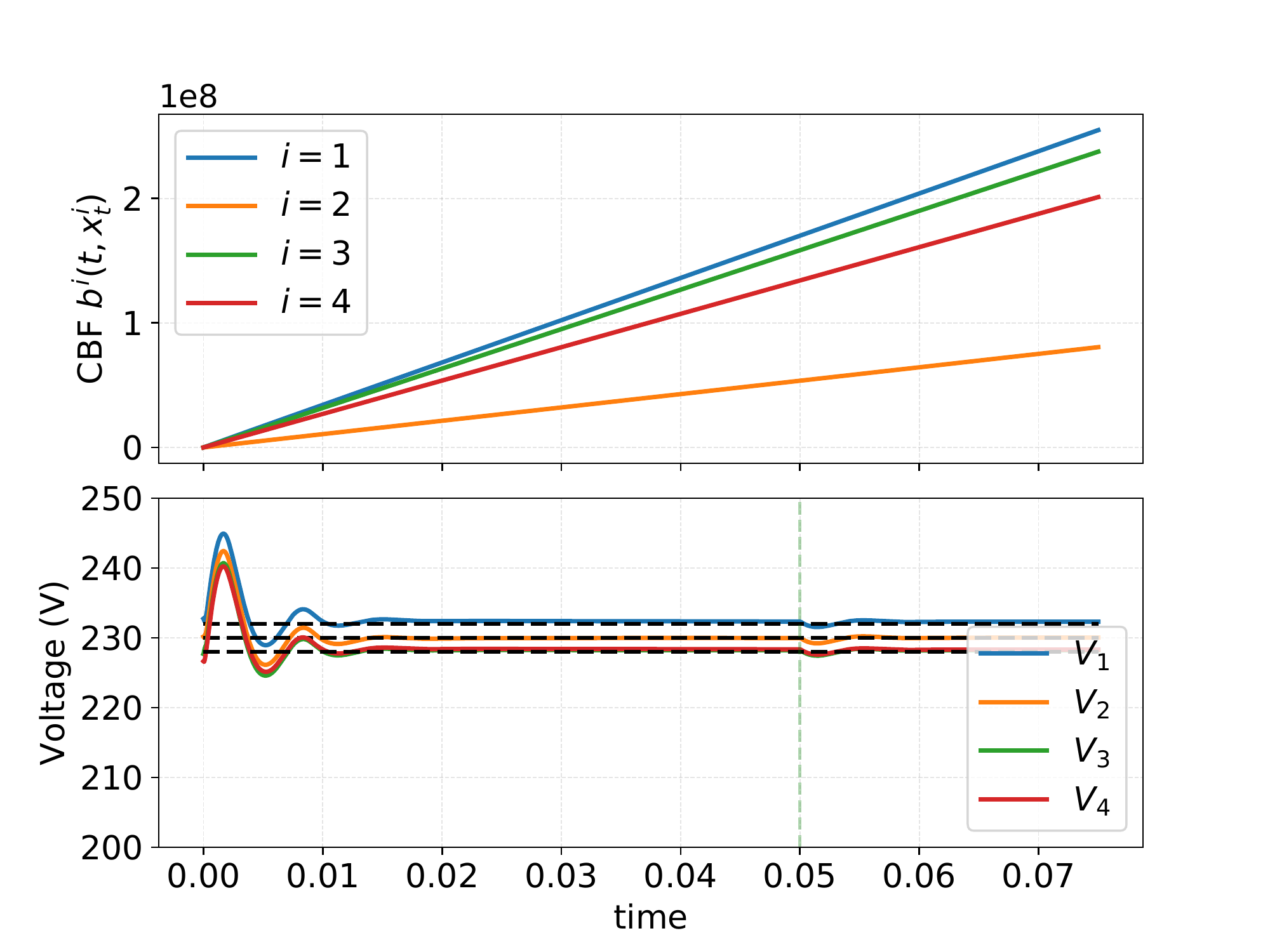}
    \caption{(Top) time evolution of control barrier function, (bottom) voltage across the load of each DGU, considering a load variation of $5\%$ at time $t=0.05$ seconds.}
    \label{fig:cbf}
\end{figure}
%
Finally, we define the instantaneous reward function at each node as
\begin{align}\label{reward}
r^i(V^i) := -k^i\left(\tilde{V^i}\right)^2
\end{align}
where $k^i \in \bR_{>0}$. 
For numerical simulation, the parameters of the microgrids are taken from~\cite[Tables~3,~and~4]{cucuzzella2019voltage}.

\begin{figure*}
    \centering
    \includegraphics[width=\textwidth]{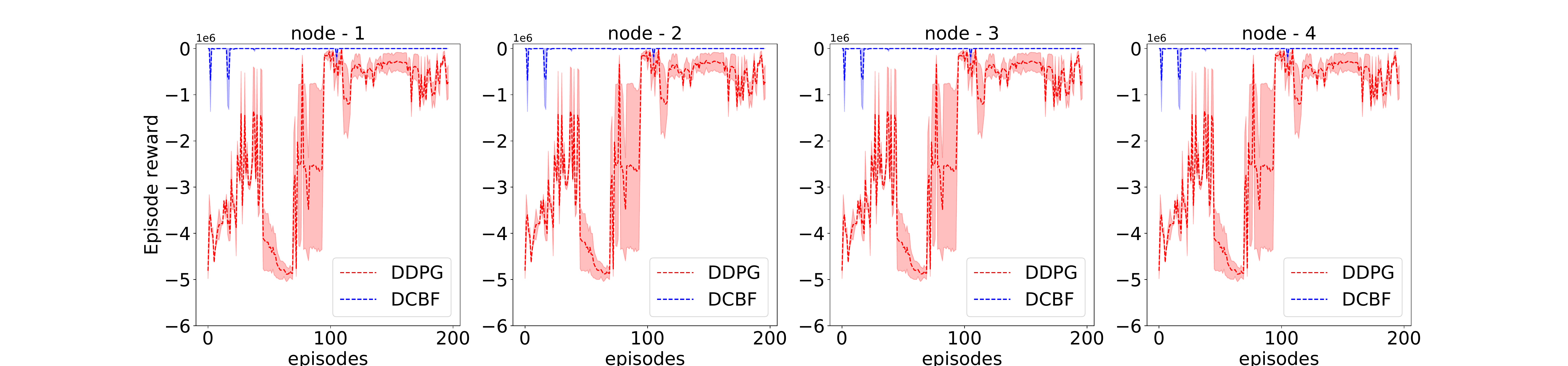}
    \caption{Comparison of accumulated rewards from nodes of DC microgrid for each episode during training using DDPG and the propose Dissipative CBF approach.}
    \label{fig:returns}
\end{figure*}

Though the general framework described in the preceding can be used with almost any RL algorithm, we chose to use Deep Deterministic Policy Gradient (DDPG) \cite{lillicrap2015continuous} to showcase the performance of Algorithm \ref{algo:DEC}.
Figure \ref{fig:returns} compares the accumulated rewards of vanilla DDPG and the proposed dissipativity-ensuring Algorithm \ref{algo:DEC} using DDPG during training. As the plot shows, Algorithm \ref{algo:DEC} coupled with DDPG converges faster that the vanilla DDPG algorithm; however, this may not be a general observation.

Next, we validate the performance of the controllers designed using the proposed approach. The voltage across the load and the value of the control barrier function at each node are plotted in Figure \ref{fig:cbf}. At $t=0$ seconds, we start by initializing the microgrid near the desired operating point. We observe that the voltage signals stabilize to their desired values. However, in the DC microgrid, the value of load $G^i$ is unknown and subject to change over time. To verify the robustness of the controller with respect to this uncertainty, the load at each DGU was increased by $5\%$ of its original value at $t=0.05$ seconds. In Figure \ref{fig:cbf} we see that, after a minor perturbation, the voltage signals again stabilized to their desired values. Furthermore, the control barrier function is positive, thus validating the dissipativity-ensuring nature of the proposed approach.


\section{Conclusions} \label{sec:conc}
In this paper, we considered the problem of designing distributed controllers to stabilize a class of networked systems, where each subsystem is dissipative. We assumed that each subsystem designs a local controller using reinforceent learning to optimize its own reward function. We develop an approach that enforces dissipativity conditions on the local controller design to guarantee stability of the entire networked system. The proposed approach was illustrated on a microgrid example.


\appendix
%
%

%
%
%
%
\subsection{Dissipativity}\label{sec:dissipativity}
Consider the following discrete time nonlinear system with state $x\in \bR^n$ and inputs $a\in \bR^m$
\begin{align}\label{sys:main}
\begin{split}
\left\{
\begin{matrix}
x_{t+1}\\
y_t
\end{matrix} 
~\begin{matrix}
=\\
=
\end{matrix}
~\begin{matrix}
f(x_t, a_t),\\
h(x_t, a_t).
\end{matrix}
\right.
\end{split}
 \end{align}
where the functions $f,~h$ as assumed to be sufficiently smooth.
Consider the mapping  $w: \bR^m, \bR^m \rightarrow \bR$. Then,  dissipativity of system $\Sigma$ with $w(a_t,y_t)$ as supply-function is defined as follows:
\begin{definition}[Dissipativity \cite{navarro2002implications}]\label{def:dissipativity}  System \eqref{sys:main} is said to be dissipative with respect to the supply function $w(a_t,y_t)$, if there exist a non-negative function $S:\bR^n \rightarrow \bR_+$, called as storage function, satisfying $S(0)=0$ such that for all $s_{t_0}\in X$, all $t> t_0\geq 0$ and all $a_t\in A$,
\begin{align}\label{diss:gen_sys_1}
    S(x_t)-S(s_{t_0}) \leq -\sum_{i=t_0}^{t-1}\mathcal D(x_t) +\sum_{i=t_0}^{t-1} w(a_t, y_t),
\end{align}
or equivalently \cite{xia2014determining},
\begin{align}\label{diss:gen_sys_2}
\sum_{i=t_0}^{t-1} w(a_t,~ y_t) \geq \sum_{i=t_0}^{t-1}\mathcal D(x_t) \geq 0
\end{align}
where $\mathcal D(x_t)\in \mathbb{R}_+$ is a non-negative function, and $s_{t}$ is the state at time $t$, resulting from state $s_{t-1}$ with input $u_{t-1}$. Furthermore, we call the system  
$QSR$ {\em dissipative} if the inequality  \eqref{diss:gen_sys_2} holds with
\begin{align}\label{diss:QSR}
\begin{split}
        w(a_t,y_t) &= -\|y_t\|_Q^2 +a_t^\top S y_t-\|a_t\|_R^2 
\end{split}
\end{align}
where $Q = Q^\top$, $S$, and $R = R^\top$ are matrices of appropriate dimensions.
\end{definition}
\begin{definition}[zero-state detectability]\label{def:zero_state_det}
Consider \eqref{sys:main} with $f(0,0)=0$, and $h(0,0)=0$. Then system \eqref{sys:main} is called zero-state detectable if 
\begin{align*}
a_t = 0~ \textit{and} ~y_t =0 ~\implies \lim_{t\rightarrow\infty}x_t \rightarrow 0.
\end{align*}
\end{definition}
%
%

%
%
\subsection{Proofs}\label{apx:proofs}

\subsubsection*{{\bf Proof of Proposition \ref{prop:cbf}}}
Without loss of generality, we assume the initial state as $s_{0} \in \rho_0$ at time $t=0$ and $b(0,s_0)\geq 0$. It suffices to show that $b(t,s_t)\geq 0$, for all $a_t\in \mathrm{DEC}(t,s_t)$. From \eqref{barrier_suff_cond:gen} and \eqref{ctrl:deterministic_pol}, for all $a_t\in \mathrm{DEC}(t,s_t)$, we have
\begin{align}\label{dyn_ineq}
    b(t+1,s_{t+1})\geq (1-\eta) b(t,s_t).
\end{align}
Now, consider the following boundary value problem:
\begin{align}\label{boundary_val_prob}
    X_{t+1} &=(1-\eta) X_t
\end{align}
with initial condition $X_{0} = b(0, s_{0})\geq 0$. Then, the solution to \eqref{boundary_val_prob} is $X_t = (1-\eta)^tX_0\geq 0$, $\forall k \in \mathbb{Z}^+$, $0<\eta\leq 1$. From \eqref{dyn_ineq} and \eqref{boundary_val_prob},
\begin{align}
    b(t,s_t) \geq  X_t.
\end{align}
Thus $\mathcal{C}$ is forward invariant.
\newline
\subsubsection*{{\bf Proof of Proposition \ref{prop:dissipativity_using_cbf}}}
Consider the barrier function $b^i(t,x^i_t)$ defined in \eqref{cbf:single_node}. 
From Proposition \ref{prop:cbf}, for all $u_t\in \mathrm{D}^i(x^i_t, \nu_t)$, it implies that $\mathcal{C}^i$ is forward invariant. 
Consequently, we have $b^i(t,x^i_t) = -\sum_{\tau =t_0}^{t-1} \tilde w \geq 0$
\begin{align}
\implies &\sum_{\tau =t_0}^{t-1} \tilde w \leq 0\\
\implies &\sum_{\tau =t_0}^{t-1}(w_n-w_d)  \leq 0\\
\implies &\sum_{\tau =t_0}^{t-1}w_d  \geq \sum_{\tau =t_0}^{t-1}w_n.
\end{align}
From Assumption \ref{ass:open_loop_dissipativity} the subsystem \eqref{sys:network:node_dynamics} is dissipative, which further implies
\begin{align}
\implies &\sum_{\tau =t_0}^{t-1}w_d  \geq \sum_{\tau =t_0}^{t-1}w_n \geq 0.
\end{align}
From Definition~\ref{def:dissipativity}, we conclude the proof.
\newline

%
%
\subsubsection*{\bf Proof of Theorem \ref{prop:stab_of_network}}
As a consequence of  Assumption \ref{ass:open_loop_dissipativity}, Proposition \ref{thm:dissipativity_RL} implies that node dynamics in closed-loop  with control input \eqref{policy:compensated} are dissipative with supply function \eqref{desired:supply-function} $w_d^i(\nu^i, y^i)$. Consequently, for all $i \in \cV$ there exist a storage function $S_d^i:\bR^n\rightarrow \bR_+$, satisfying
\begin{align}\label{proof:node_diss}
    S_d^i(x^i_t) \leq S_d^i(x^i_{t_0}) + \sum_{t=t_0}^{t-1}w_d^i(\nu^i, y^i).
\end{align}
From Assumption \ref{ass:open_loop_dissipativity_edge}, the edge dynamics are dissipative with supply-function $w_e^k(\mu^k, \omega^k)$. Consequently, for all $k \in \left\{1, \dots, M\right\}$, there exist a storage function $S_e^i:\bR^m\rightarrow \bR_+$, satisfying
\begin{align}\label{proof:edge_diss}
    S_e^k(z^k_t) \leq S_e^i(z^k_{t_0}) + \sum_{t=t_0}^{t-1}w^k_e(\mu_t^k, \omega_t^k).
\end{align}
Consider $S(s_t) = \sum_{i=1}^{N}S_d^i(x^i_t) + \sum_{k=1}^{M}S_e^k(z^k_t)$, consequently
\begin{subequations}
\begin{align}
    &S(s_t) -S(s_{t_0}) \label{proof:stability_of_network:eq1a}\\
    \leq & \sum_{i=1}^{N}\sum_{t=t_0}^{t-1}w_d^i(\nu^i, y^i) + \sum_{k=1}^{M}\sum_{t=t_0}^{t-1}w^k_e(\mu_t^k, \omega_t^k)\nonumber\\
     = & \sum_{t=t_0}^{t-1}\sum_{i=1}^{N} w_d^i(\nu^i, y^i) + \sum_{t=t_0}^{t-1}\sum_{k=1}^{M}w^k_e(\mu_t^k, \omega_t^k)\nonumber\\
    \leq&\sum_{t=t_0}^{t-1}\left(\nu^{\top}S_{\nu}^\top y_{\nu} - \alpha\|\nu\|^2_2 -\beta\|y\|^2_2 +\mu^{\top}S_{e}^\top \omega - \delta_e\|\mu\|_2^2 \right.\nonumber\\
       &\left.- \epsilon_e\|\omega\|_2^2\right)\label{proof:stability_of_network:eq1b}\\
       \leq&\sum_{t=t_0}^{t-1}\left(\omega^{\top}\mathcal{B}^\top S_{\nu}^\top y_{\nu} - \alpha\|\mathcal{B}\omega\|^2_2 -\beta\|y_{\nu}\|^2_2-\beta\|y_{u}\|^2_2 \right.\nonumber\\
       &\left.-y_{\nu}^{\top}\mathcal{B}^\top S_{e}^\top \omega - \delta_e\|\mathcal{B}y_{\nu}\|_2^2 - \epsilon_e\|\omega\|_2^2\right)\label{proof:stability_of_network:eq1c}\\
       %
       %
            =&-\sum_{t=t_0}^{t-1}\left(\|\omega\|^2_{\mathcal{B}_{\delta}(\alpha)} +\|y_{\nu}\|^2_{\mathcal{B}_{\epsilon}(\beta)}+\beta\|y_{u}\|^2_2\right)\label{proof:stability_of_network:eq1d}
\end{align}
\end{subequations}
In \eqref{proof:stability_of_network:eq1a} we use \eqref{proof:node_diss} and \eqref{proof:edge_diss}. 
In \eqref{proof:stability_of_network:eq1b} we use the interconnection laws from \eqref{sys:graph:inter_laws_gen}. In \eqref{proof:stability_of_network:eq1c}, we use Assumption \ref{ass:S}. This implies that the overall networked system is stable.

Furthermore, consider $\mathcal{B}_{\delta}(\alpha) > 0$, and $\mathcal{B}_{\epsilon}(\beta)> 0$. Then from \eqref{proof:stability_of_network:eq1d} there exists a forward invariant set $\Pi$ and by LaSalle's invariance principle, 
the solutions that start in $\Pi$ converge to the largest invariant set contained in
\begin{align}
    \Pi \cap \left\{s\in\bR^{n+p}|~\omega = 0,~ y=0\right\}.
\end{align}
Moreover, from \eqref{sys:graph:inter_laws_gen} this implies $\mu =0 ,~\nu=0$. 
From Assumption \ref{ass:open_loop_dissipativity} and $R_u> 0$ this further implies that $u=0$. Finally on this set, we have $(y=0, u=0, \nu=0)$ and $(\omega=0, \mu=0)$. Given that that subsystems \eqref{sys:network:node_dynamics} and \eqref{sys:network:edge-dynamics} are  zero-state detectable, the trajectories in
$\Pi$ converges asymptotically to the largest invariant set contained in
\begin{align}
    \Pi \cap \left\{s=0\right\},
\end{align}
following~\cite[Corollary~ 4.2.2]{l2gain}.

%
%
\bibliographystyle{IEEEtran}
\bibliography{ref}
\end{document}